\pdfoutput=1

\documentclass[aps,prb,reprint,superscriptaddress,showkeys]{revtex4-1}

\usepackage{graphics,amsmath}
\usepackage{graphicx}

\begin{document}

\title{Direct time domain sampling of sub-THz coherent acoustic phonon spectra in SrTiO$_3$ using ultrafast X-ray diffraction}

\keywords      {Ultrafast X-ray diffraction, coherent phonons, SrRuO$_3$/SrTiO$_3$, pump-probe spectroscopy}

\author{Roman Shayduk}
  \email{roman.shayduk@helmholtz-berlin.de}
  \affiliation{Helmholtz-Zentrum Berlin f\"ur Materialien und Energie GmbH, Wilhelm-Conrad-R\"ontgen Campus, BESSY II, Albert-Einstein-Str. 15, 12489 Berlin Germany}

\author{Marc Herzog}
\affiliation{Institut f\"ur Physik und Astronomie, Universit\"at Potsdam, Karl-Liebknecht-Str. 24-25, 14476 Potsdam, Germany}

\author{Andre Bojahr}
\affiliation{Institut f\"ur Physik und Astronomie, Universit\"at Potsdam, Karl-Liebknecht-Str. 24-25, 14476 Potsdam, Germany}

\author{Daniel Schick}
  \affiliation{Institut f\"ur Physik und Astronomie, Universit\"at Potsdam, Karl-Liebknecht-Str. 24-25, 14476 Potsdam, Germany}

\author{Peter Gaal}
  \affiliation{Helmholtz-Zentrum Berlin f\"ur Materialien und Energie GmbH, Wilhelm-Conrad-R\"ontgen Campus, BESSY II, Albert-Einstein-Str. 15, 12489 Berlin Germany}

\author{Wolfram Leitenberger}
  \affiliation{Institut f\"ur Physik und Astronomie, Universit\"at Potsdam, Karl-Liebknecht-Str. 24-25, 14476 Potsdam, Germany}

\author{Hengameh Navirian}
  \affiliation{Institut f\"ur Physik und Astronomie, Universit\"at Potsdam, Karl-Liebknecht-Str. 24-25, 14476 Potsdam, Germany}

\author{Mathias Sander}
  \affiliation{Institut f\"ur Physik und Astronomie, Universit\"at Potsdam, Karl-Liebknecht-Str. 24-25, 14476 Potsdam, Germany}

\author{Jevgenij Goldshteyn}
  \affiliation{Helmholtz-Zentrum Berlin f\"ur Materialien und Energie GmbH, Wilhelm-Conrad-R\"ontgen Campus, BESSY II, Albert-Einstein-Str. 15, 12489 Berlin Germany}

\author{Ionela Vrejoiu}
  \affiliation{Max-Planck-Institut f\"ur Mikrostrukturphysik,
Weinberg 2, D-06120 Halle, Germany}

\author{Matias Bargheer}
\affiliation{Institut f\"ur Physik und Astronomie, Universit\"at Potsdam, Karl-Liebknecht-Str. 24-25, 14476 Potsdam, Germany}
\affiliation{Helmholtz-Zentrum Berlin f\"ur Materialien und Energie GmbH, Wilhelm-Conrad-R\"ontgen Campus, BESSY II, Albert-Einstein-Str. 15, 12489 Berlin Germany} 

\begin{abstract}
 We synthesize sub-THz longitudinal quasi-monochromatic acoustic phonons in a SrTiO$_3$ single crystal using a SrRuO$_3$/SrTiO$_3$ superlattice as an optical-acoustic transducer. The generated acoustic phonon spectrum is determined using ultrafast X-ray diffraction. The analysis of the generated phonon spectrum in the time domain reveals a k-vector dependent phonon lifetime. It is observed that even at sub-THz frequencies the phonon lifetime agrees with the 1/$\omega^2$ power law known from Akhiezer's model for hyper sound attenuation. The observed shift of the synthesized spectrum to the higher $q$ is discussed in the framework of non-linear effects appearing due to the high amplitude of the synthesized phonons.
\end{abstract}

\maketitle

\section{Introduction}

The increasing importance of coherent phonon spectroscopy in material science is related to the growing problem of heat dissipation in modern nanoscale devices. This problem is impossible to solve without detailed understanding of underlaying phonon-phonon and phonon-electron interactions on the nanoscale. One of the methods to study this processes is coherent phonon spectroscopy, in which a particular phonon spectrum is excited coherently in the sample and detected optically. Research efforts in this direction resulted in significant progress in generation and detection of coherent phonons in various materials. The available phonon frequency has reached the THz acoustic limit \cite{wu2007a} and basically the whole phonon frequency range nowadays could be excited coherently. However, convenient optical detection methods based on Raman\cite{rama1928c,cho1990a} or Brillouin\cite{Chia1964} scattering allow for the observation of phonons excited only in the vicinity of the Brillouin zone center. Therefore, sub-THz acoustic phonons could be accessed optically only in multilayer structures, in which the acoustic dispersion branch backfolds many times inside a mini-Brillouin zone of a multilayer \cite{colv1980a}. Modern progress in pulsed laser techniques as well as in multilayer fabrication has lead to a set of successful experiments in which the coherent zone-folded superlattice phonons have been optically excited and detected \cite{colv1980a,colv1985a,bart1999}. However, these optical methods are insensitive to the THz frequency phonons which have propagated into the bulk of the crystal due to the unfolding of the phonon dispersion curve. Convenient optical methods based on Brillouin scattering in this case have a detection limit in the 100\,GHz range given by the wave vector magnitude of the optical light \cite{briv2011a,Kais1966a}. Recently, ultrafast X-ray diffraction (UXRD) has become available to extend the accessible phonon frequency range to above 100\,GHz. It has been used successfully to study both the time-domain structure of optically excited zone-folded coherent acoustic phonons in epitaxial multilayers\cite{barg2004b} as well as to observe the propagation of unfolded phonons into the bulk \cite{Trig2008a}.

In this paper we report our new UXRD experiments from coherent quasi-monochromatic longitudinal acoustic phonons in SrTiO$_3$ synthesized by fs-laser excitation of SrRuO$_3$/SrTiO$_3$ (SRO/STO) epitaxial multilayers. Using UXRD we determine the laser excited phonon spectrum in SrTiO$_3$ and monitor the modification of the spectrum in the time domain. The epitaxial multilayers were prepared using Pulsed Laser Deposition \cite{vrej2008a}. The experiments are carried out at the BESSY EDR beamline using a unique setup for a 1\,MHz repetition rate UXRD experiments.

The experiments are done in a traditional scheme which uses an optical delay line to change the time interval between the optical pump and the X-ray probe pulses. We use infrared optical pulses with the wavelength of 1.03\,$\mu$m for pumping and 8\,keV X-rays for probing the lattice dynamics. The important feature of this setup is the simultaneous acquisition of the X-ray photons scattered from the sample before and after the pumping optical pulse. This makes the X-ray intensity difference signal sensitive only to those changes in the crystal lattice which were exclusively initiated by the optical pulses. For a further details we refer to a recent publication describing the setup\cite{navi2012a}.

\section{Theory}

\subsection{Synthesis of quasi-monochromatic coherent acoustic phonons}

\begin{figure}[tb]
  \includegraphics[width=.45\textwidth]{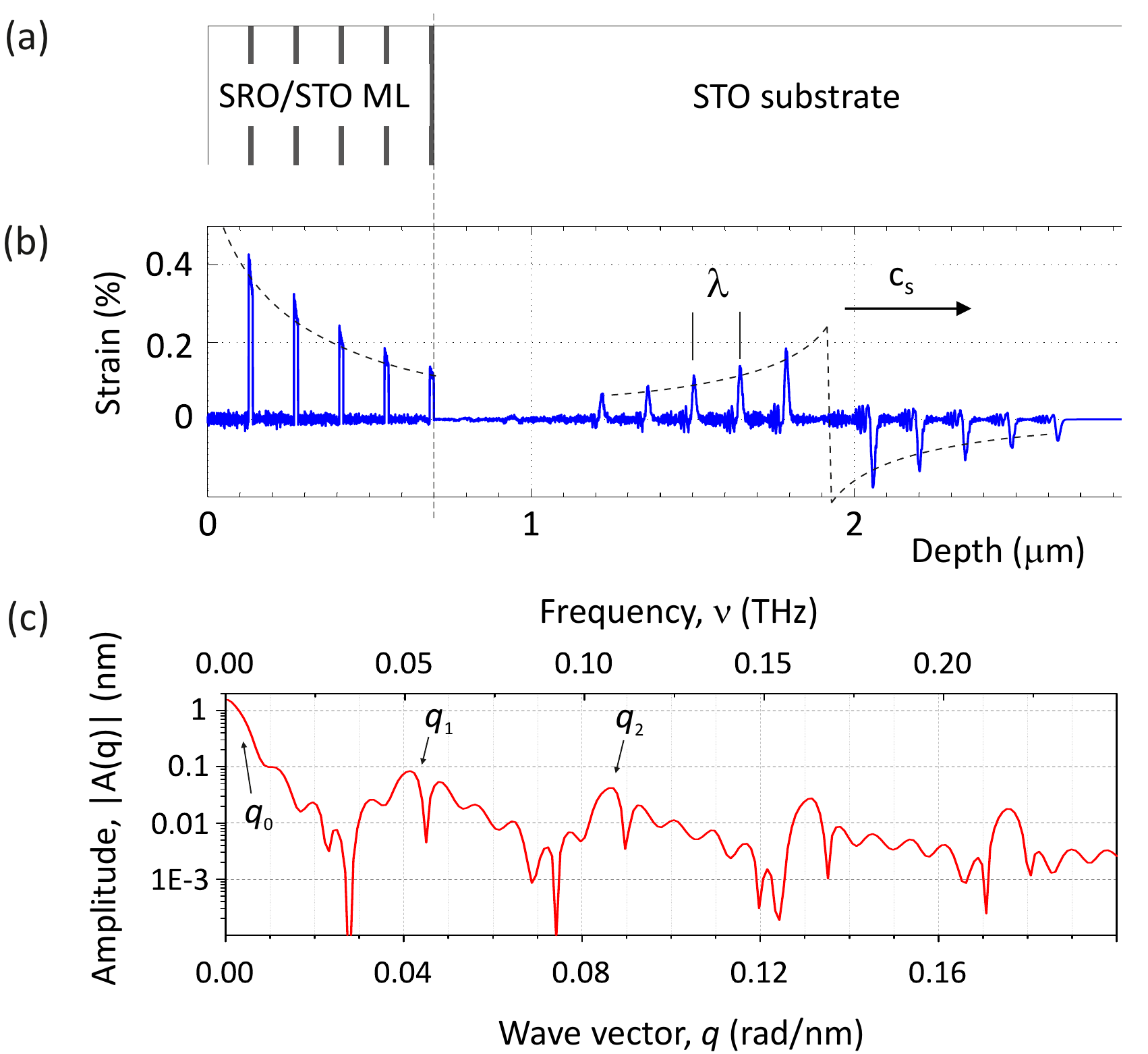}
  \caption{(Color online) (a) Sketch of the 5-period epitaxial SrRuO$_3$/SrTiO$_3$ multilayer on top of the SrTiO$_3$ substrate. (b) Calculated instant strain profile in the sample along the surface normal, taken at 250\,ps after the multilayer excitation. (c) Calculated spectrum of the strain pulse in the wave vector and frequency domains induced by the optical excitation. The vertical scale stands for the excitation fluence of 6\,mJ/cm$^2$ used in the experiment.}
  \label{PhononSimulations}
\end{figure}

Recent studies showed that the optical excitation of a metal transducer by a sequence of ultrashort laser pulses is an efficient method to generate sub-THz quasi-monochromatic longitudinal acoustic (LA) phonons \cite{choi2005,klie2011a,herz2012c}. In essence, the repetitive generation of bipolar strain pulses by the laser-excited transducer \cite{thom1986a,schi2012b} forms a phonon wavepacket of narrow spectral bandwidth propagating throughout the substrate. In this report we consider a different approach which uses a spatial repetition instead of a temporal one, i.e., the excitation of a periodic metal-dielectric multilayer (superlattice) with a single ultrashort laser pulse (see Fig.~\ref{PhononSimulations}(a)).
This way the so-called superlattice phonon mode is excited \cite{barg2004b,herz2010a,herz2012a,herz2012b,boja2012a} which subsequently unfolds into the substrate thereby forming LA phonon wavepackets with similarly narrow spectral bandwidth \cite{Trig2008a,herz2012b}. In most cases the laser-induced coherent lattice dynamics may be calculated using either a model of a continuous elastic medium \cite{thom1986a} or a linear-chain model (LCM) of masses and springs \cite{herz2012b}. Here we employ the latter approach which will be more appropriate for short-period superlattices and automatically accounts for the acoustic phonon dispersion. The details of the numerical model can be found in Ref.~\citenum{herz2012b}. The excellent agreement of such a linear-chain model with related optical and UXRD experiments has been previously demonstrated for various sample structures \cite{herz2012a,herz2012b,herz2012c,boja2012a,boja2012b,gaal2012a,schi2012b}.

Further we consider linear chain calculations for a five-period SRO/STO multilayer with the spatial period $D=140\,$nm which is schematically shown in Fig.~\ref{PhononSimulations}(a). The structural parameters of the sample which are determined by static high-resolution x-ray diffraction (HRXRD) \cite{shay2011a} are collected in Table~\ref{SampleParametersTable}. In addition, the longitudinal sound velocities of the individual materials are shown\cite{bell1963}. The results of the linear-chain calculations using the parameters given in Table~\ref{SampleParametersTable} and the experimental pump fluence of 6~mJ/cm$^2$ are shown in the Figs.~\ref{PhononSimulations}(b) and (c).

The graph in Fig.~\ref{PhononSimulations}(b) shows the calculated one-dimensional strain profile in the sample 250~ps after the excitation. Figure~\ref{PhononSimulations}(c) plots the spectral amplitude of the linear-chain eigenvectors (normal modes) as a function of the eigenfrequency. This amplitude spectrum is solely determined by the initial conditions \cite{herz2012b}. For the bulk STO substrate the eigenvectors are plane elastic waves with wave vector $q$ satisfying the well-known dispersion relation of acoustic phonons \cite{ashc1976a}. As Fig.~\ref{PhononSimulations}(b) illustrates, the optical excitation of the metal layers of the multilayer system results in the generation of a coherent strain wavepacket propagating into the STO substrate at the longitudinal sound velocity \cite{herz2012b}. The resulting wavepacket inside the STO substrate attains the particular shape shown in Fig.~\ref{PhononSimulations}(b), namely, five leading compression pulses and five trailing expansion pulses which are separated by $\lambda \approx 140\,$nm, respectively. In other words, the metal/dielectric multilayer acts as the photoacoustic transducer synthesizing the coherent LA phonon wavepacket in the STO substrate. The sharp static profile of the thermal strain inside the multilayer (see Fig.~\ref{PhononSimulations}(b)) remains unchanged with time because the linear chain model neglects the effect of heat diffusion. The heat diffusion in multilayers is a complicated separate topic which lays out of the scope of this article. In this article we focus on the coherent lattice dynamics in the STO substrate which occur at a later timescale when the effect of heat diffusion within the multilayer does not play a role. For the detailed description of the wavepacket strain profile and its generation we refer to our earlier works \cite{herz2012b}.

\begin{table}[tb]
\caption{Structural and mechanical properties employed in the calculations.}
\vspace{3mm}
\begin{tabular}{l|l|l|l}
Material & Lattice constant & Thickness & Sound speed \\
\hline
Substrate STO  & 3.905 \AA & 10\,$\mu$m& 7.9 nm/ps \\
STO in ML & 3.92 \AA & 127\,nm & 7.8 nm/ps \\
SRO in ML & 3.95 \AA & 13\,nm & 6.3 nm/ps \\
\end{tabular}
\label{SampleParametersTable}
\end{table}

The calculated amplitude spectrum in Fig.~\ref{PhononSimulations}(c) contains several equidistant peaks. The most pronounced peak at $q_0=0$\,rad/nm is responsible for the overall bipolar shape of the wavepacket \cite{thom1986a,herz2012b,schi2012b}. The width of the peak is deterined by the total thickness of the multilayer ($\Delta q\approx\pi/5\lambda$). The peak around $q_1=2\pi/\lambda \approx 0.046$\,rad/nm corresponds to the characteristic spatial period $\lambda$ of the wavepacket. The non-sinusoidal shape of the wavepacket gives rise to the higher harmonics at integer multiples of $q_1$.

Altogether, we find that using a periodic metal-dielectric multilayer as photoacoustic transducer we can generate LA phonon wavepackets similar to the quasi-monochromatic wavepackets produced by multiple-pulse excitation of a thin metal film \cite{choi2005,klie2011a,herz2012c}. In both cases the wavepackets exhibit narrow spectral bandwidth and higher harmonics of lower amplitude.

\subsection{Ultrafast X-ray diffraction from sub-THz elastic waves}

The X-ray diffraction from crystals which are subject to a strong acoustic field is a well-established topic \cite{enti1988a,saue1998a,Zolo2002}. However most of the previous studies deal with strain fields generated by surface acoustic wave (SAW) transducers. Such devices normally generate acoustic waves with wavelengths longer than either the X-ray extinction length or the X-ray coherence length. The X-ray diffraction from a crystal lattice perturbed by such waves results in modifications of the Bragg peak shape within the Darwin width or in the appearance of diffuse scattering contributions in the vicinity of the peak \cite{Sand1995,Zolo1998a}. The description of the X-ray scattering from such waves usually requires dynamical X-ray diffraction theory.

Here we consider quasi-monochromatic coherent LA phonons which in fact are elastic waves at hypersonic frequencies. The corresponding wavevectors $q$ are large enough to allow for coherent Bragg-like scattering of X-rays from the associated ''moving gratings''. Due to the sufficiently high $q$ vectors of the quasi-monochromatic phonons the X-ray scattering contributes in the off-Bragg region in which the X-ray scattering efficiency from the bulk of the crystal is small. This allows to probe the X-rays exclusively scattered from the wavepacked with only minor perturbation by the bulk-scattered X-ray wavefield.

In this section we introduce the necessary theoretical basis which allows a thorough interpretation of the UXRD experiments. Since we intend to apply the following theory to study the laser-induced structural dynamics in one dimension we restrict ourselves to a one-dimensional formulation.

A plain elastic wave with wave vector $q$ contributes to the scattered X-ray intensity if the X-ray scattering vector $Q$ is given by
\begin{equation}
  Q = G \pm  q
\label{ScatteringFromPhonon}
\end{equation}
where $G$ is a reciprocal lattice vector and Q = $| \mathbf{k}_{}$ - $\mathbf{k}_i |$ is the X-ray scattering vector \cite{Lind2000,Reis2001,Lars2002,herz2012c}.

From simulations of the scattered X-ray intensity using dynamical theory of X-ray diffraction one finds that the X-ray intensity scattered from the crystal perturbed by a bunch of elastic waves can be well described by the equation
\begin{equation}
  \langle I_p(Q) \rangle_t = I_{up}(Q) + \alpha A(q)^2
\label{IntensityFromPhonon}
\end{equation}
where $I_p(Q)$ and $I_{up}(Q)$ is the scattered X-ray intensity from the perturbed and unperturbed crystal, respectively. The angle brackets stand for time averaging. The function $A(q)$ is the spectral amplitude of the elastic wave with wave vector magnitude $q=|Q-G|$ and $\alpha$ is some constant. It is worth to show here that formula (\ref{IntensityFromPhonon}) is equivalent to the expression describing thermal diffuse scattering (TDS) from acoustic phonons \cite{Warr1990}. To show this we need to relate the energy of a classical plane elastic wave in the crystal with the phonon population. The energy of plane elastic waves in the classical linear theory of elasticity is proportional to the squared product of the wave amplitude $A$ and frequency $\omega$
\begin{equation}
  E(\omega) \propto A^2\omega^2
\label{ClassicalEnergy}
\end{equation}
In a crystal lattice this corresponds to the energy of the corresponding vibrational normal mode which is associated with a single harmonic oscillator. According to quantum mechanics the energy of a harmonic oscillator with angular frequency $\omega$ is
\begin{equation}
E(\omega)= \hbar \omega (n+\frac{1}{2})
\label{QuantumMechanicalEnergy}
\end{equation}
where $n$ is the excitation level. That is, the energy of the vibrational normal modes is quantized and $n$ refers to the number of phonons in the crystal having the angular frequency $\omega$. Therefore, the following relationship between the excited classical amplitude spectrum of elastic waves and the phonon population holds
\begin{equation}
A(q_i)^2\propto(n(\omega_i)+\frac{1}{2})/\omega_i\approx\frac{n(\omega_i)}{\omega_i}
\label{PhononPopulation}
\end{equation} in which index $i$ identifies the normal mode.
The combination of (\ref{IntensityFromPhonon}) and (\ref{PhononPopulation}) yields
\begin{equation}
\langle I_p(Q) \rangle_t-I_{up}(Q)\propto\frac{n(Q-G)}{\omega(Q-G)}
\label{PhononPopulationFromIntensity}
\end{equation}
which is a one-dimensional equivalent of the relation for TDS derived by Warren\cite{Warr1990}.

We thus conclude that UXRD from a quasi-monochromatic strain pulse directly measures the squared spectral amplitudes of the plane elastic waves constituting the strain pulse. As an example we consider the reciprocal lattice vector $G_{002}$ of STO and rewrite (\ref{IntensityFromPhonon}) and (\ref{PhononPopulationFromIntensity}) into
\begin{eqnarray}
  A(q) & \propto & \sqrt{\langle I_p(G_{002}+q) \rangle_t-I_{up}(G_{002}+q)}  \label{AmplitudeFromIntensity} \\
  n(q) & \propto & \omega(q)\big(\langle I_p(G_{002}+q) \rangle_t-I_{up}(G_{002}+q)\big)  \label{AmplitudeFromIntensity2}
\end{eqnarray}

In the standard $\theta$-2$\theta$ geometry applied in our UXRD experiments, the magnitude of the phonon wave vector $q$ is:
\begin{equation}
  q = \frac{4\pi}{\lambda_X}|\sin \theta - \sin \theta_0|
\label{PhononVectorFromAngle}
\end{equation}
where $\lambda_X$ is the X-ray wavelength, $\theta$ is the X-ray incidence angle with respect to the sample surface ((001) crystallographic plane) and $\theta_0$ is the Bragg angle.

\begin{figure}[tp]
  \includegraphics[width=.45\textwidth]{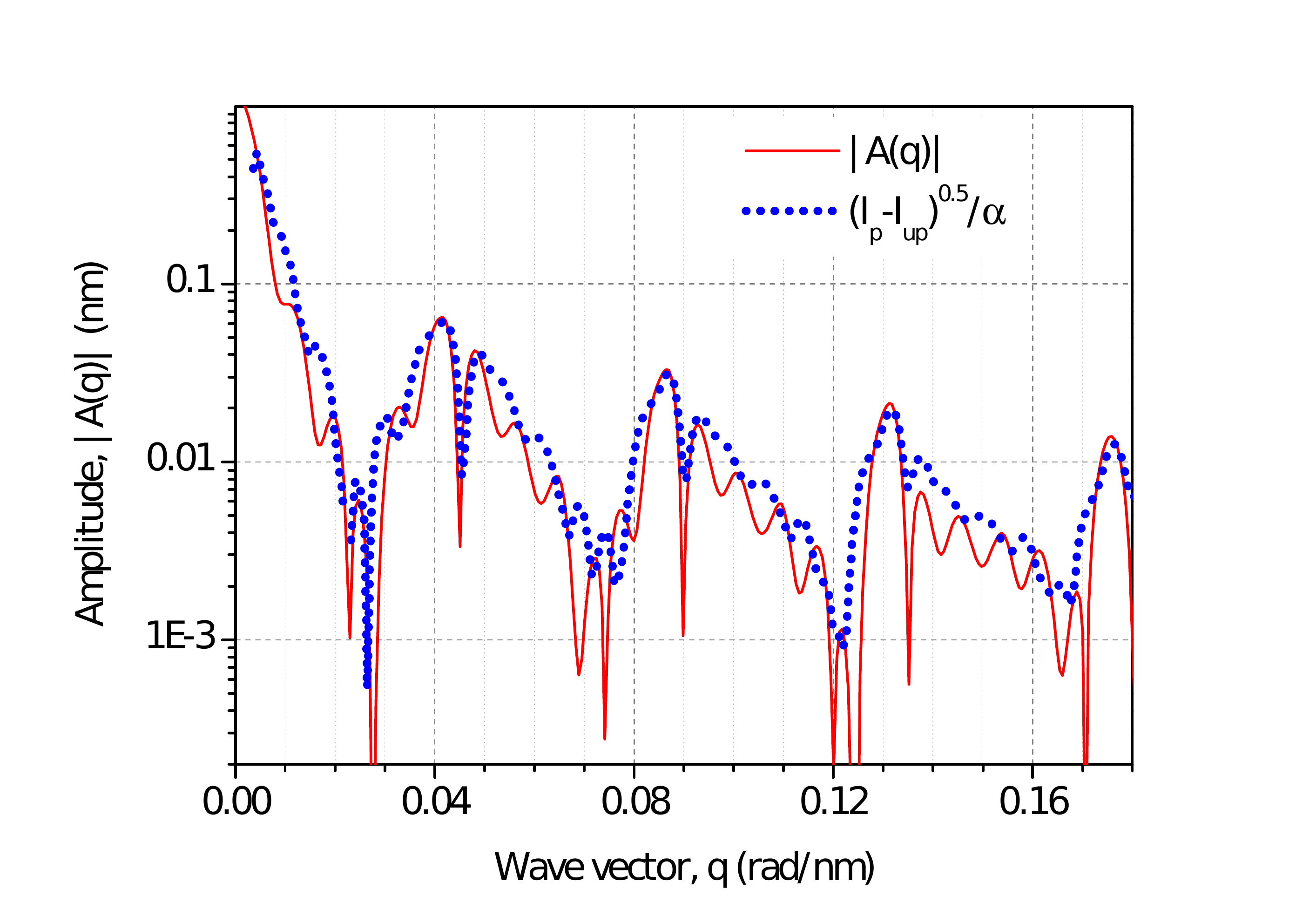}
  \caption{(Color online) Comparison of the calculations of the spectral strain amplitudes of phonons with the corresponding x-ray intensity difference signal of perturbated-unperturbated structures. The red solid line indicates the calculated phonon spectrum the sample in the $q$ domain. The values in the vertical axis correspond to the phonon spectral amplitudes calculated for the excitation fluence of 6~mJ/cm$^2$. The blue bullets show the square root of the x-ray intensity difference signal calculated for the perturbated and unperturbated structures. The vertical scale factor for the UXRD signal is arbitrary. }
  \label{XRDtheory}
\end{figure}

To demonstrate, that (\ref{AmplitudeFromIntensity}) is applicable to our case we compare the calculated amplitude spectrum of the laser-excited strain waves to the dynamical UXRD simulations from the same acoustically perturbed sample \cite{herz2012a}. The calculated spectrum for the laser fluence of 6~mJ/cm$^2$ is plotted in Fig.~\ref{XRDtheory} as a red solid line. The dynamical UXRD calculations are performed for 200 time steps within the interval from 100\,ps to 300\,ps after the excitation and then time-averaged. The blue symbols in the Fig.~\ref{XRDtheory} show the scaled time averaged square root of the intensity differences [cf. (\ref{AmplitudeFromIntensity})] obtained from the dynamical UXRD calculations.

We see that the curves almost coincide although the fine structure of the UXRD-related curve (blue symbols) slightly deviates from the actual spectrum (red solid line). This is due to the fact that the scattered X-ray intensity from a propagating strain wavepacket actually oscillates in the time domain at each fixed $q$-vector with the frequency of the corresponding phonon mode. These oscillations were successfully observed in pioneering experiments with the advent of UXRD \cite{Lind2000,Reis2001,Lars2002}. The classical explanation for the oscillations is the interference of X-rays scattered from a moving grating (elastic wave) and the X-rays scattered from the static component of the crystal lattice.  This is an X-ray analog of Brillouin oscillations in all-optical experiments\cite{boja2012c}. In our case, as we can see, this interference is not very strong. Therefore, the shape of the blue curve slightly depends on the averaging time window. To eliminate these artifacts, the averaging time window should be either much longer than any phonon vibration period or we need to fit an integer number of vibrations for each $q$. The averaging over many vibrational periods is not possible in our case, because the actual phonon life time is only several vibration periods as we will see later.

To finish this section we briefly review the conditions at which the approximation (\ref{AmplitudeFromIntensity}) should be valid:

\begin{itemize}
  \item
    The interatomic displacement in the strain wave is much less than the interatomic distance:
    \begin{equation}
      |r_m-r_n|\ll_{m\neq n}|a(m-n)|
      \label{ValidityConditions1}
    \end{equation} in which $a$ is the interatomic distance, $m$ and $n$ are the index number of atoms. This is required by the perturbation theory of X-rays scattered from a dynamical lattice\cite{Warr1990}.
  \item
    The wave vector of a phonon is much smaller than any reciprocal lattice vector:
    \begin{equation}
      \mathrm{q}\ll\mathrm{G}
      \label{ValidityConditions2}
    \end{equation} This is necessary to avoid the signal overlap from the adjacent Bragg reflections of the crystal
  \item
    The wavelength of a phonon mode is much smaller than both the X-ray coherence and the X-ray extinction lengths:
    \begin{equation}
      \mathrm{q}\gg\frac{1}{l}
      \label{ValidityConditions3}
    \end{equation} in which $l$ is either X-ray coherence or X-ray extinction lengths, depending on which one is larger.
  \item
    The X-ray intensity is time-averaged over many vibrational periods
\end{itemize}

\section{Experimental results}

We performed UXRD experiments on the laser-excited five-period SRO/STO multilayer with the structure parameters presented in the Table \ref{SampleParametersTable}. In this section we present the experimental results which evidence the presence of a propagating quasi-monochromatic LA phonon wavepacket. We discuss the dynamics of the first and second-order transient diffraction peaks and the corresponding dynamics of the strain pulse.

In the experiment we acquire the X-ray photons scattered from the sample 50\,ns before each optical pulse and at a given probe delay after each optical pulse. The corresponding scattered X-ray intensities from perturbated and unperturbated sample are thus defined as $I_p$ and $I_{up}$, respectively. The time resolution of the experiments was 100\,ps due to the limited X-ray pulse length, therefore the measured X-ray intensity is time averaged over multiple phonon vibrations.

\begin{figure}[tp]
  \includegraphics[width=.48\textwidth]{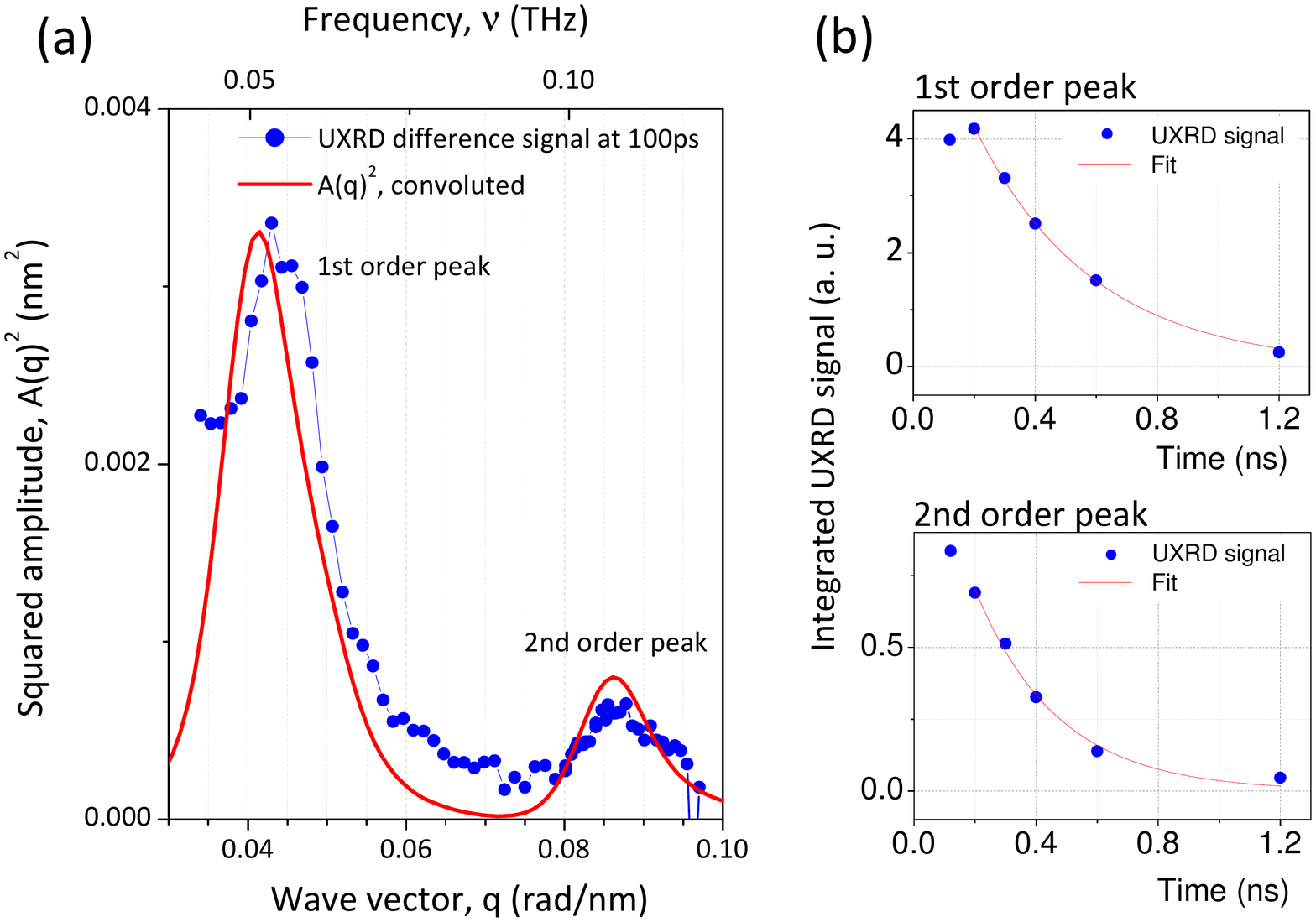}
  \caption{(Color online) (a) The solid dots show the UXRD difference signal with an arbitrary vertical scale factor. Red is the calculated spectrum of the synthesized wavepacket. (b) The integrated UXRD difference signal of the first two order phonon peaks as a function of time. The solid lines indicate the exponential fit. }
  \label{PhononDecayTime}
\end{figure}

Figure~\ref{PhononDecayTime}(a) shows the measured x-ray intensity difference signal $I_p-I_{up}$ (blue symbols) in the vicinity of the STO (002) Bragg peak (not visible) at 100~ps after the laser-pulse excitation. The experimental incidence angle $\theta$ has been converted into the phonon wavevector using (\ref{PhononVectorFromAngle}). The vertical scale for the measured data is arbitrary.
The UXRD data exhibit the first and second-order spectral components of the synthesized quasi-monochromatic phonon wavepacket inside the STO substrate at wave vectors $q_1 \approx 0.045$~rad/nm and $q_2 \approx 0.09$~rad/nm, respectively. Given the longitudinal sound velocity in STO (cf. Table~\ref{SampleParametersTable}), the linear phonon dispersion relation of acoustic phonons implies the corresponding hypersonic frequencies $\nu_1 \approx 55$~GHz and $\nu_2 \approx 110$~GHz. The non-vanishing contributions between the phonon peaks are due to the diffraction from the laser-heated multilayer. However, since the lattice constants throughout the multilayer are larger than that of the substrate (cf. Table~\ref{SampleParametersTable}), the X-ray scattering from the multilayer is rather weak in this angular range. The red solid line in Fig.~\ref{PhononDecayTime}(a) shows the squared amplitude spectrum of the propagating sound wave as obtained from the linear-chain model. The shown spectrum includes the convolution with a Gaussian resolution function having a full width at half maximum (FWHM) of $15\cdot10^{-3}$\,nm$^{-1}$ to fit the angular resolution of the UXRD experiment. The main contribution to the XRD peak broadening is due to the sample bending according to the stationary laser heat load\cite{navi2012a}. The UXRD signal shows very good agreement with the convoluted spectrum in terms of position, relative intensity and width of the first and second-order phonon peaks. This verifies the relation between the measured x-ray intensity and the amplitude spectrum of the coherent strain wave derived in (\ref{AmplitudeFromIntensity}).

In the following we discuss the intensity changes of the measured phonon peaks with time. During the first 100~ps after laser-excitation the intensity of the phonon peaks builds up \cite{herz2012c} due to the unfolding of the initially excited superlattice phonon mode into the substrate \cite{Trig2008a,herz2012b}. Subsequently, the integrated intensity of the phonon peaks decays exponentially as is evidenced by the blue symbols in Fig.~\ref{PhononDecayTime}(b). The red solid lines show fits according to the function
\begin{equation}
  f(t)=\Delta I_0 e^{-\frac{t}{\tau_{\mathrm{exp}}}}
  \label{FittingFunction}
\end{equation}
where the two fitting parameters $\Delta I_0$ and $\tau_{\mathrm{exp}}$ are the amplitude and decay time of the measured signal. The data points before 200~ps after the excitation were excluded from the fit since the wavepacket may not yet fully propagated from the multilayer to the substrate. The extracted decay times $\tau_{\mathrm{exp}}$ and standard deviations $\sigma_{\mathrm{exp}}$ are shown in Table~\ref{table}.
\begin{table}[tb]
\caption{Comparison of the experimentally observed UXRD intensity decay time $\tau_{\mathrm{exp}}$, the apparent decay time due to x-ray absorption $\tau_{\mathrm{abs}}$ and the derived phonon lifetime $\tau_{\mathrm{ph}}$ for the first and second-order phonon peaks. The corresponding standard deviations $\sigma_{\mathrm{exp/abs}}$ are also shown.}
\vspace{3mm}
\begin{tabular}{c|c|c|c}
$q$, rad/nm & $\tau_{abs}$, ps & $\tau_{exp}$, ps & $\tau_{ph}$, ns\\
\hline
0.045 & 450 & 373 $\pm$ 12 & 2.2 $\pm$ 0.5\\
0.09 & 450 & 235 $\pm$ 30 & 0.49 $\pm$ 0.13\\
\end{tabular}
\label{table}
\end{table}

There are two major reasons for the observed decrease of the UXRD peak intensities. First, the absorption of the X-rays by the crystal reduce the sensitivity of the X-rays to the wavepacket as it propagates deeper into the substrate. Second, the dissipation of energy from the elastic wave due to the finite phonon lifetimes leads to a decay of the strain amplitude.

Considering the first reason, the decay time of the UXRD signal exclusively due to X-ray absorption is related to the X-ray absorption coefficient $\mu=0.056\,\mu m^{-1}$ by
\begin{equation}
  \frac{1}{\tau_{\mathrm{abs}}} = \frac{2\mu c_s}{\sin\theta}  = \frac{8\pi \mu c_s }{Q\lambda_X}
\end{equation}
where $c_s=7.9\,nm/ps$ is the longitudinal sound speed in the substrate\cite{bell1963}. The relative variation of the x-ray scattering vector $Q$ during the presented UXRD experiments is $10^{-3}$ which implies that $\tau_{\mathrm{abs}}$ is virtually independent of the observed phonon wavevectors $q$. In the measured off-Bragg region the X-ray extinction due to dynamical X-ray diffraction is negligible, therefore only the angular independent X-ray absorption is relevant. Under the chosen experimental conditions we estimate a signal decay time of $\tau_{\mathrm{abs}} \approx 466$~ps due to the x-ray absorption. Nevertheless, since this value is critical for the correct interpretation of the experimental data, we have performed dynamical XRD calculations based on results of the linear-chain lattice dynamics in harmonic approximation which excludes the effect of phonon damping\cite{herz2012c,herz2012b}. The simulations yield the $q$-independent value of 450$\pm$5\,ps for the decay constant due to X-ray absorption which we will use in the following.

Regarding the second reason for the decay of the UXRD phonon signals, we assume an exponential law for the decrease of the phonon population $n(q,t)$ and define the associated decay time $\tau_{\mathrm{ph}}$. According to (\ref{AmplitudeFromIntensity}) the corresponding intensity of the scattered X-rays possesses the same decay constant.

Therefore, the UXRD signal decay mechanisms introduce the relationship between the phonon lifetimes $\tau_{\mathrm{ph}}$, X-ray absorption time constant $\tau_{\mathrm{abs}}$ and the experimentally measured time constant $\tau_{\mathrm{exp}}$:
\begin{equation}
\frac{1}{\tau_{\mathrm{exp}}}=\frac{1}{\tau_{\mathrm{abs}}}+\frac{1}{\tau_{\mathrm{ph}}}
\label{DecayTimesRelationship}
\end{equation}

The phonon lifetimes $\tau_{\mathrm{ph}}$ extracted from the UXRD experiments according to (\ref{DecayTimesRelationship}) are given in the Table~(\ref{table}) for the wave vector magnitudes corresponding to the first and second order phonon peaks. The standard deviation $\sigma_{\mathrm{ph}}$ for the phonon lifetimes are calculated from the standard deviations $\sigma_{\mathrm{exp}}$ of the experimental time constants according to the error propagation relation.

\begin{figure}
  \includegraphics[width=.5\textwidth]{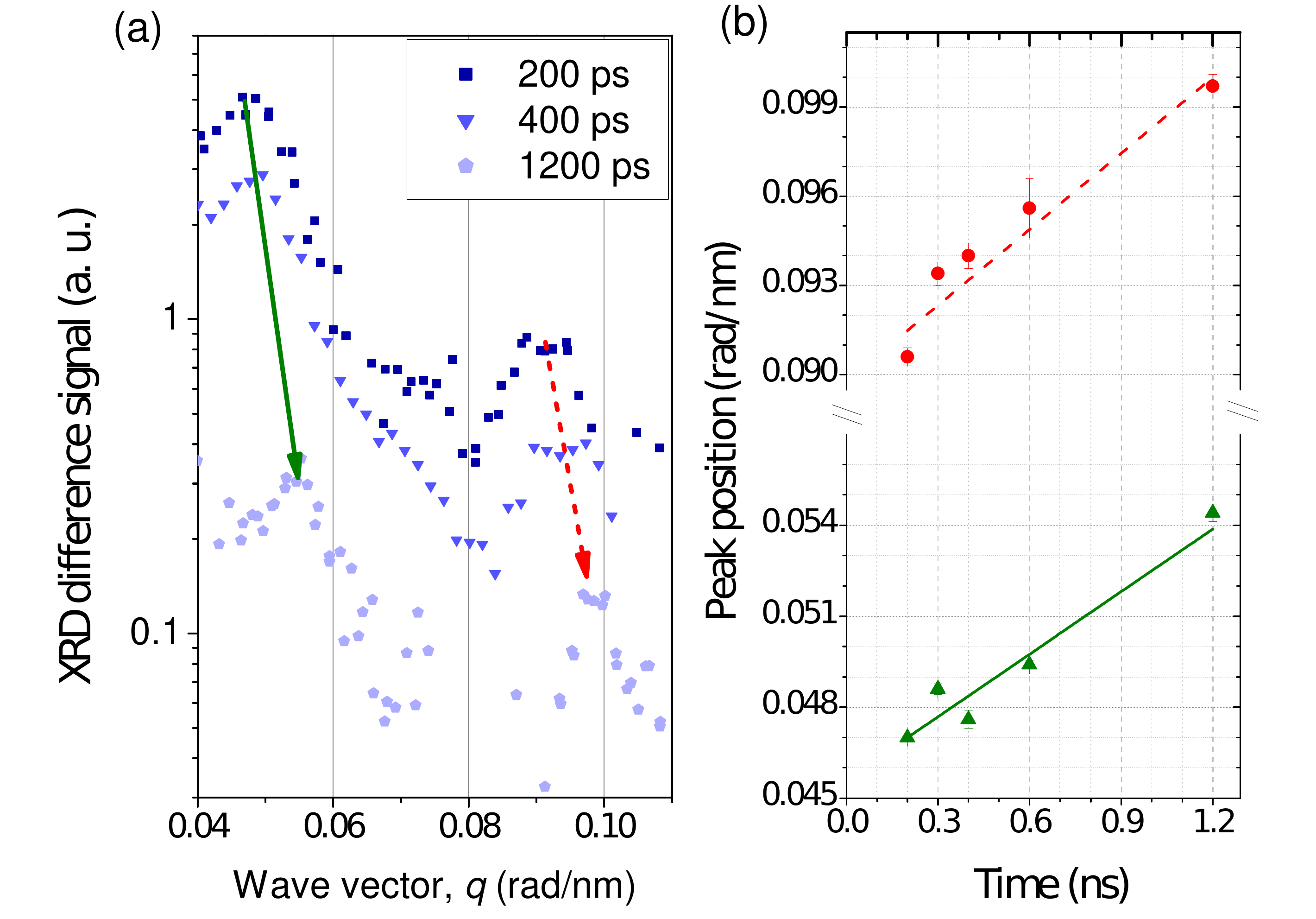}
  \caption{(Color online) (a) Transient first and second-order phonon diffraction peaks at various probe delays. The intensity decay due to sound attenuation is accompanied by a continuous shift to larger wave vectors as is indicated by the arrows. (b) First and second-order peak position determined by Gaussian fits as a function of time delay. The solid lines show the linear fits to the experimental points.}
  \label{PhononPeakShift}
\end{figure}

 The important observation is the fact that the determined lifetimes differ by a factor of approx. 4 whereas the related phonon wave vector differ by a factor of 2. In addition to the signal decay we see a gradual drift of the phonon peak position to higher values of $q$ as the wavepacket propagates. This is seen from the Fig. ~\ref{PhononPeakShift}(a) in which the phonon spectrum is shown for different time delays.

In the following section we discuss the physical interpretation of the described observations.

\section{Discussion}

The central observation from the transient phonon-induced diffraction peaks presented in the previous section is the apparent quadratic decrease of the phonon lifetime with phonon wave vector $q$. This observation is in agreement with the $1/\omega^2$ law predicted by the Akhiezer's sound attenuation mechanism \cite{akhi1939a}.

There are basically two theories which explain the attenuation of a hypersonic waves in dielectric crystals due to incoherent anharmonic phonon-phonon scattering. These are the Landau-Rumer theory \cite{land1937a} and the Akhiezer theory \cite{akhi1939a} which have different application limits:
\begin{eqnarray}
\omega&\ll& \frac{1}{\tau_{th}} \qquad \textrm{Akhiezer} \label{Akhiezer} \\
\omega&\gg& \frac{1}{\tau_{th}} \qquad \textrm{Landau-Rumer} \label{Rumer}
\end{eqnarray}
where $\omega$ is the angular frequency of the hypersonic wave and $\tau_{th}$ is the mean thermal phonon relaxation time. That is, in both cases the elastic energy of hypersound waves decreases with time due to the interaction with incoherent thermal phonons. The mean thermal phonon relaxation time  $\tau_{th}$ can be estimated from the thermal conductivity $k$, heat capacity $C_V$, average sound speed $v_s$ and mass density $\rho$ by the relation \cite{ashc1976a}
\begin{equation}
k = \frac{1}{3}\rho C_V v_s^2 \tau_{th}
\label{ThermalConduct}
\end{equation}
For STO at room temperature (\ref{ThermalConduct}) yields $\tau_{th} \approx 0.26$~ps. The experiments are performed at room temperature. More precisely, the sample persisted at around 400\,K during the actual measurements due to the thermal load from the laser\cite{navi2012a}. Condition (\ref{Akhiezer}) is fulfilled for this temperature range. That is, for the presented experiments Akhiezer's theory of relaxation damping could be applied, hereby explaining the observed ratio of the phonon lifetimes for the first and second-order phonon diffraction peaks. However the UXRD data exhibit additional features which cannot be explained within Akhiezer's sound attenuation model. We extracted the transient peak positions by Gaussian fits to the data shown in Fig. ~\ref{PhononPeakShift}(a) and the results are plotted as symbols in Fig.~\ref{PhononPeakShift}(b). The solid lines indicate linear fits to the phonon peak positions as a function of time. Clearly, a gradual shift of the first and second-order spectral components to higher $q$-values can be observed as the phonon wavepacket propagates deeper into the STO substrate.

A recent study revealed the nonlinear propagation of large-amplitude sound wavepackets in STO at room temperature \cite{boja2012b}. In addition to the Akhiezer-like attenuation of the coherent LA phonons the authors observed transient changes of the acoustic spectrum due to coherent anharmonic phonon-phonon scattering within the wavepacket. The lattice anharmonicity gave rise to a strain-dependent longitudinal sound velocity. In particular, the sound velocity of compressive (tensile) parts of the wavepacket was found to increase (decrease) with the strain amplitude. This effect led to an anomalous dispersion of the wavepacket and the corresponding modification the phonon spectrum.

Accordingly, we expect the first compressive half of the wavepacket shown in Fig.~\ref{PhononSimulations}(b) to propagate faster than the second tensile half. Moreover, the individual pulses inside the respective parts also exhibit different velocities due to the exponential amplitude distribution determined by the optical penetration of the pump light in SRO. For both the compressive and tensile parts of the strain pulse the spatial separation $\lambda$ of the individual pulses of the wavepacket is reduced as it propagates, i. e. the wavelength of both sub-packets is decreased. This explains the observed shift of the phonon peaks to larger $q$-values.

The presented UXRD data thus evidence the influence of two different effects on the propagation of LA phonon wavepackets generated by periodic multilayers. First, the inevitable attenuation of the wavepackets by Akhiezer's relaxation damping and, second, the change of spatial and spectral shape of the wavepacket by nonlinear sound propagation. Both effects influence the observed phonon lifetime, however, for a quantitative determination of the respective contributions additional measurements have to be performed. The results of our earlier all-optical experiments having much stronger excitation were successfully explained solely in the framework of nonlinear acoustics\cite{boja2012b}. We believe that at the presented experimental conditions the influences of both damping mechanisms the non-linear acoustic propagation and the Akhiezer's relaxation are comparable.

UXRD has the advantage of measuring the lattice dynamics directly and quantitatively, i.e. the absolute amplitude of the lattice motion is determined. The wavevector range over which acoustic phonons in bulk material are accessible is very large. In particular resolving the second order phonon peak as presented in this paper or higher orders is possible. The extension of the UXRD detection of acoustic phonons in amorphous materials is challenging the available x-ray fluence, since the Bragg spots are dispersed in diffraction rings.

On the other hand all-optical picosecond acoustics \cite{thom1986a,cho1990a,Chia1964,bart1999,briv2011a}, in principle, do not require crystalline materials and for transparent media, the propagation of strain pulses can be monitored over longer distances. With current technology femtosecond time resolution is standard in all-optical experiments, while it is still a challenge in x-ray technology, which essentially resolved by free-electron lasers. High time-resolution permits the determination of the wave vector dependent sound velocity in addition to the damping time.

We believe that the UXRD-based methods and the all-optical methods do not compete with each other but complete each other, together providing a more complete picture of the complex coherent phonon dynamics for a broader range of frequencies and wavevectors and for a broader class of materials and experimental conditions.

\section{Conclusions}

This report presents ultrafast x-ray diffraction (UXRD) studies on laser-excited periodic SrRuO$_3$/SrTiO$_3$ multilayers which are epitaxially grown on a SrTiO$_3$ substrate. The ultrafast heating of the metallic SrRuO$_3$ layers by ultrashort laser pulses generates coherent longitudinal acoustic phonons which eventually propagate into the substrate as a quasi-monochromatic coherent LA phonon wavepacket at hypersonic frequencies. We discussed the properties of such wavepackets in detail and derive equations which show that UXRD is a powerful tool to measure the spectral phonon population and its dynamics. The presented UXRD data evidence the formation of quasi-monochromatic coherent phonon wavepacket. We extract the phonon lifetimes of the first and second-order peaks of the phonon spectrum. The observed quadratic decrease of the phonon lifetime with increasing phonon wavevector $q$ is in accordance with Akhiezer's mechanism of relaxation damping. Shifts of the peaks corresponding to the excited phonons to larger q-values are interpreted as a modification of the spatial shape profile due to the non-linear wave propagation leading to a strain dependent sound velocity. This considerably modifies the observed phonon lifetimes.

In essence, UXRD provides a detailed and direct view on the complex nonlinear evolution of phonon-wavepackets, including incoherent damping of the phonon amplitude by coupling to other modes and specific coherent changes of the wavevector spectrum.

\section{Acknowledgements}

We thank the BMBF for funding via 05K 2012-OXIDE.


%

\end{document}